\documentstyle[preprint,aps,prl]{revtex}
\begin{document}
\draft

\title{
The de Haas-van Alphen effect in canonical and grand canonical
multiband Fermi liquid}
\author{ A.S. Alexandrov$^1$ and A.M. Bratkovsky$^2$ }
\address{
$^1$~Department of Physics, Loughborough University of Technology,
 Loughborough, Leics. LE11 3TU, U.K.\\
$^2$~Department of Materials, University of Oxford, Oxford OX1 3PH, U.K.}
\date{August 4, 1995}
\maketitle

\begin{abstract}

A qualitatively different character of dHvA  oscillations has been found
in a multiband (quasi)two dimensional Fermi liquid
 with a fixed fermion density $n_{e}$ (canonical ensemble) compared with
an open system where the  chemical potential $\mu$ is kept fixed (grand
 canonical ensemble). A new fundamental period $P_{f}$ appears
when $n_{e}$ is fixed, a damping of
the Landau levels is relatively small and a background density of states
is negligible.
$P_{f}$ is determined by the {\em total} density rather than by the partial
densities of carriers in different bands: $P_{f}=1/(2n_{e}\phi)$ for
spin-split Landau levels and $P_{f}=1/(n_{e}\phi)$ in the case of spin
degenerate levels where $\phi$ is the flux quantum.
\end{abstract}

The de Haas-van Alphen
 oscillations of  susceptibility as a function of the inverse
magnetic field $1/H$ are well studied both experimentally and
theoretically \cite{David84}. The frequency of the oscillations is
proportional to the extremal cross section of the Fermi surface.
Therefore, in the multiband metals one would expect several  different
dHvA periods corresponding to the independent contributions of
different bands.\cite{David84,Lan39,LifKos55}
The dHvA effect in closed and open systems was analysed by Dingle back
in  1951 \cite{Din51} with the conclusion that there
is practically no difference because the  dependence of the
chemical potential on the magnetic field, in the case of fixed $n_{e}$,
is very weak.
In this Letter we show  that while this is true  for three dimensions
and(or) for the  relatively large
damping of Landau levels, the dHvA effect is qualitatively
different   in the near two-dimensional (2D)
canonical Fermi liquid compared with  the grand canonical ensemble
 if the damping is relatively weak.

 If we  keep the total
 number of electrons per area, $n_{e}$, in the near 2D metal fixed
 the chemical potential  will oscillate with inverse magnetic field,
$1/H$. These oscillations are responsible for a new fundamental period in
the two- or multi-band Fermi liquid. The value of the period can be
evaluated by using a simple dHvA resonance condition. There are no
partially occupied Landau levels at the resonance.
 In   two dimensions each of the Landau levels  can be occupied by $pH$
 carriers per $cm^{2}$, where  $p=1/(2\phi)$ if the levels are spin split,
 and $p=1/\phi$ if they are not ($\phi=\pi \hbar c/e$ is
the flux quantum).
Then the  dHvA resonance
 condition for fixed $n_{e}$ is
 \begin{equation}
 {e(H+\Delta H)(N-1)\over{2\pi\hbar c}}=n_{e},
\end{equation}
where $N=1,2,3...$ is determined by
\begin{equation}
{eH N\over{2\pi\hbar c}}=n_{e}.
\end{equation}
Combining Eq.(1) and Eq.(2) we obtain the fundamental dHvA period
\begin{equation}
P_f={p\over n_e},
\label{eq:pfund}
\end{equation}
which is independent of the band structure according to the
following simple argument.
 In a {\em multiband} metal  the Landau levels
(LL) will be occupied sequentially depending on their  energy.
The number of electrons which the LL can accommodate is always equal to $pH$,
where $p$ is constant [$=e/(2\pi\hbar c)$], i.e it does not depend on
 the characteristics of the band. It means that the population of the
LLs in multiband metals with field  in the  canonical ensemble
 is equivalent to that of a  one-band metal.
The individual bands will then, generally, show up in an intensity
of peaks in susceptibility
resulting in additional oscillations of the moment and susceptibility
superimposed on the fundamental one.
However, the main Fourier
component will be the fundamental one with frequency $1/P_{f}$.
On the other hand, in the open system this Fourier component is absent or
significantly suppressed compared with the individual band oscillations.
The conditions for the observation of the fundamental frequency seem
to be only marginally stricter, due to its relatively higher value,
than those for the ordinary dHvA effect,
and it should be observable in near 2D electrically insulated specimens.

 It turns out that it is quite difficult to obtain the
fundamental frequency in multiband metals
with the use of the standard Poisson summation formula.
In what follows, we shall therefore consider firstly the case of a
clean 2D metal at zero temperature
with (i) the total number of electrons being fixed and (ii) the chemical
potential $\mu$ being fixed (i.e. the metal is well connected to
 some  ``reservoir'' of electrons).
Then we shall estimate the effect of the broadening of the Landau levels.

Let us consider a two-band 2D metal
with different band masses, where the bands are split
into series of Landau levels (Fig.~1),
\begin{equation}
\epsilon_i(n) = \Delta_i + \hbar\omega_i(n + \frac{1}{2}),~~~n~=~0, 1, ...
\label{eq:LL}
\end{equation}
where $i=1, 2$ is the band index, and $\omega_i=eH/(m_ic)$ is the
cyclotron frequency.
Each level is degenerate, contains $pH$ states,
and is broadened by collisions with impurities into a Lorentzian with
Dingle width $\sim \hbar/\tau$.\cite{Din51} We shall assume that
$\hbar/\tau \ll \hbar \omega$, and describe the  situation where
the $N_1$ levels in the first band and $N_2$ levels in the second band are
occupied, and the last Landau level  in the first band is partially
occupied by $xpH$ electrons, where $x=n_e/(pH)-[n_e/(pH)]$, $0<x<1$,
and $[a]$ stands for the integer part of $a$.

Let us now consider canonical and grand  canonical ensembles.\\

\noindent
(i) Canonical ensemble ($n_e=const$).
Generally, the orbital moment is found from
\begin{equation}
M = -\left( \partial F\over \partial H\right)_{ T,V}=
-{2k_BT\over\pi}\int_{-\infty}^\infty d\epsilon
{\rm Im~Tr}{\partial G(\epsilon,H)\over \partial H}
\ln\left( 1 + e^{ {\mu-\epsilon}\over T } \right),
\end{equation}
where $G$ is the electron Green's function which accounts for
collisions,\cite{YuB61} and the chemical potential $\mu$
is defined by the conservation of the total number of electrons, $n_e$.
In the multiband case for a clean metal it would amount to a rather
complicated
 non-linear equation if we were to apply the standard
Poisson summation formula.
To elucidate the physics, we shall consider first the limiting
case of zero temperature in the clean limit.
By counting the number of electrons in the Landau levels we obtain
\begin{equation}
(N_1-1)pH + xpH + N_2pH = n_e,
\end{equation}
with a similar relation when the LL in the second band is partially
occupied.
Then we immediately have for the period of the dHvA oscillations
$P_f =  {1/H} - {1/(H+\Delta H)} = {p/n_e}$,
 i.e. the same fundamental period as we have found before,
Eq.~(\ref{eq:pfund}),
which is {\em the same for all bands}.
For the energy we have, if the partially occupied LL belongs to
the first band,
\begin {equation}
E = \sum_{n_1=0}^{N_1-1}pH\epsilon_1(n_1) +xpH\epsilon_1(N_1) +
\sum_{n_2=0}^{N_2}pH\epsilon_2(n_2),
\label{eq:e}
\end{equation}
with a similar equation when the partially occupied LL belongs to
the second band;
the moment is found from $M=-dE/dH$. We are interested in
the semiclassical regime, where the total number of occupied
Landau levels is large, $N_{tot}=n_e/(pH) \gg 1$, as well as the
number of occupied LL in each band,
\begin{equation}
N_i = \left[ {m_ic\over eH}(\mu-\Delta_i) -\frac{1}{2} \right] \gg 1.
\label{eq:Ni}
\end{equation}
The chemical potential is pinned to the partially occupied LL
and oscillates about the {\em mean} value
$
\bar{\mu}~=~({n_e\hbar/c + \sum_i m_i\Delta_i})/\sum_i m_i,
$
which is field independent.\\

\noindent
(ii) Grand canonical ensemble, $\mu=const$.
In that case the period is defined by the condition of the LL crossing
the fermi level,
$ \mu= \Delta_i + {e\hbar H\over m_ic}(N_i +\frac{1}{2}) $,
and we find that the oscillations have independent periods
\begin{equation}
P_i = {e\over m_ic} {1\over \mu-\Delta_i},
\label{eq:Pi}
\end{equation}
with the ratio
\begin{equation}
{P_i\over P_j} = {m_j(\mu - \Delta_j) \over m_i(\mu - \Delta_i)} =
{S_i^{-1}\over S_j^{-1}},
\label{eq:ps}
\end{equation}
where $S_i$ is the area of Landau orbit in a plane perpendicular to the
field. We see that if there were no band offset, the ratio of the periods
would have been given by the ratio of inverse masses.
If the Dingle temperature is much smaller than the inter-level spacing,
the total energy will be given
by Eq.~(\ref{eq:e}) without the term containing $x$.
It is interesting to note the existence of a simple {\em sum rule}
relating individual and fundamental dHvA frequencies
\begin{equation}
\sum_i {1\over P_i} = {1\over P_f}
\label{eq:sumrule}
\end{equation}

The moments calculated for two cases with $m_1:m_2=1:4$, and
$\Delta_2-\Delta_1=0.333$ are presented in Fig.~2. It is  seen
that there is
a vast difference between the two regimes. In grand canonical ensemble
($\mu=const$) the total moment is a sum of two periodic contributions coming
with different periods, whereas in the canonical ensemble the moments in the
bands follow the fundamental period and change in antiphase following
the progressive occupation of the LLs with decreasing field.
Although the fundamental period mirrors the level occupation
$x$, the resulting total moment shows a very complicated behavior
which reflects the individual periods with the ratio (\ref{eq:ps}).
The reason for this irregularity of the field dependence of the moment
(kinks occuring when the current LL is only partially filled)
is the {\em crossing} of the Landau levels belonging to different bands
which can only occur when simultaneously $\Delta_i \neq \Delta_j$
and $m_i \neq m_j$.
This is because the energy levels in the light band move faster with
the field and occasionally the partially occupied LL in the light band
sinks below the
highest occupied LL in the heavy band and leaves it only partially occupied.
This results in discontinuous changes in the moment and susceptibility.
The total moment in canonical ensemble oscillates about zero [Fig.~2(a)],
similarly to
a  one-band 2D metal.\cite{David84}  In grand canonical ensemble there is a
steady flow of electron into the system with reducing field resulting in
overall monotonous change of the absolute value of the moment [Fig.~2(b)].

The Fourier transform of the total moment  (Fig.~3) in the
canonical ensemble shows the rich structure
of the Fourier components. The light (L) band shows up at low frequency
with an intensive second harmonic (L2) and
has a weight much larger compared to the heavy (H) band which gives
a signal at higher frequency. The fundamental period (F)
has the largest weight
and intensive second (F2) and third (F3) harmonics.
It is easily seen that our {\em sum rule}
(\ref{eq:sumrule}) holds: L+H=F.

It is important to consider a situation when the LLs in one band
(second, for certainty) are strongly smeared out. Then
if the Dingle temperature\cite{Din51} for the second band is bigger than the
LL separation in the first band, $\hbar/\tau > \hbar\omega_1$,
the second band
can be viewed as a uniform background density of states,
$\nu_2(E)\approx const$,
for the series of the LL in the first band, $\nu_1(E)=
pH\sum_n \delta(E-E_n)$, where $\nu_i(E)$ stands for the density of states.
One can easily write down the equation for $\mu$ in this case and
apply the Poisson
formula to sum over the occupied LLs. This equation can then be solved by
successive iterations with the result that pinning of the chemical potential
at the partially occupied LL vanishes and only the {\em standard}
one-band period, $P=p/n_1$, remains.
The reason for this is
that the second band would work as a ``reservoir'' of electrons for
the first band and canonical and grand canonical ensembles would become
indistinguishable. The same is true if there is a significant background
in the density of states.
It sets  standard constraints for the observation of the fundamental
period: $max[k_BT,(\hbar/\tau_i)] < \hbar \omega_{min}$,
which could be met in clean samples at low temperatures.
We have also performed the calculation for a more realistic model of
Landau levels with finite width $\Gamma_i$ and obtained similar results.
The fundamental period is seen in systems obeying this condition and
dissapears with increasing width $\Gamma_i$. In three dimensions each LL
develops into a band in such a way that the density of states has a smooth
background weakly depending on energy ($\propto \sqrt{E}$).
%This background plays an analogous role to the damping in the multiband 2D
%system.
Consequently, the difference
between the two ensembles disappears with increasing dimensionality.
We expect, therefore, a  new fundamental period $P_{f}$ to appear
when $n_{e}$ is fixed, the damping of
the Landau levels is relatively small and the background density of states
is negligible.

The difference between canonical and grand canonical ensembles could be
relevant for results of recent dHvA measurements on near 2D
Sr$_2$RuO$_4$.\cite{mac}
Three
dHvA frequencies of 3.05 kT, 12.7 kT, and 18.5 kT with the ratio being
surprisingly close to  1:4:6 have been found.  If they are the genuine
frequencies
corresponding to three bands as given by band calculations,
the fundamental frequency is expected to be 34.25~kT.
If, on the other hand, we consider only the first two frequencies
as being genuine, the fundamental
one would be at 15.75 kT, which is somewhat smaller than the third
measured frequency of 18.5 kT. It would be interesting therefore to go into
the region of about 35~kT to look for the fundamental dHvA
frequency in Sr$_2$RuO$_4$, and also to compare the dHvA effect
in electrically insulated and noninsulated samples.

We appreciate enlightening discussions with N.F. Mott,
D. Shoenberg, D. Khmelnitskii, G. Lonzarich, and
 A. Mackenzie. We thank the Materials Modelling Laboratory at
 Oxford University for
the provision of computer facilities.

\begin{figure}
\caption{
The schematic representation of the
electronic structure of multiband near 2D metal.
The two bands corresponding to heavy and light carriers
with the offset $\Delta$ are shown. In an external magnetic field
the bands are split up into a series of the Landau levels whose
population depends on a position of the chemical potential $\mu$.
}
\end{figure}

\begin{figure}
\caption{
(a) The dHvA oscillations in canonical ensemble with two undamped
bands, $m_1:m_2=1:4$ and $\Delta_2  - \Delta_1=0.333$, at the
surface density $n_e=1$. Arrows indicate the points of Landau levels
crossing (see text). Bold line: total moment, dotted and dashed lines:
partial band contributions.
(b) The dHvA oscillations in the grand canonical ensemble with the
same parameters. Bold line: total moment, dotted line: light band,
and dashed line: heavy band.
In the top panel the filling
fraction $x$ of the partially occupied Landau levels is shown.
}
\end{figure}

\begin{figure}
\caption{
The Fourier transform of the moment in (a) canonical ensemble with
clearly resolved components of the light band (L and second harmonic L2),
the heavy band (H), and the fundamental period and its higher harmonics
(F, F2, and F3, respectively). The sum rule for the
dHvA frequencies holds: L+H=F. (b) Grand  canonical ensemble. Only the
standard individual harmonics (L and H) are clearly seen.
}
\end{figure}


\begin{references}

\bibitem{David84} D. Shoenberg, {\em Magnetic Oscillations}
(Cambridge Univ. Press, Cambridge, 1984).
\bibitem{Lan39} L.D. Landau, in: D. Shoenberg,
 Proc.Roy.Soc. {\bf A170},
341 (1939).
\bibitem{LifKos55} I.M. Lifschitz and A.M. Kosevich, Zh. Eksp. Teor.
Fiz. {\bf 29}, 730 (1955).
\bibitem{Din51} R.B. Dingle, Proc. Roy. Soc. {\bf A211}, 500,517 (1951).
\bibitem{YuB61} Yu. A. Bychkov, Sov. Phys. JETP {\bf 12}, 977 (1961).
\bibitem{mac} A.P. Mackenzie, S.R. Julian, A.J. Diver, G.G. Lonzarich,
Y. Maeno, S. Nishizaki, and T. Fujita, submitted to
Nature.

\end{references}
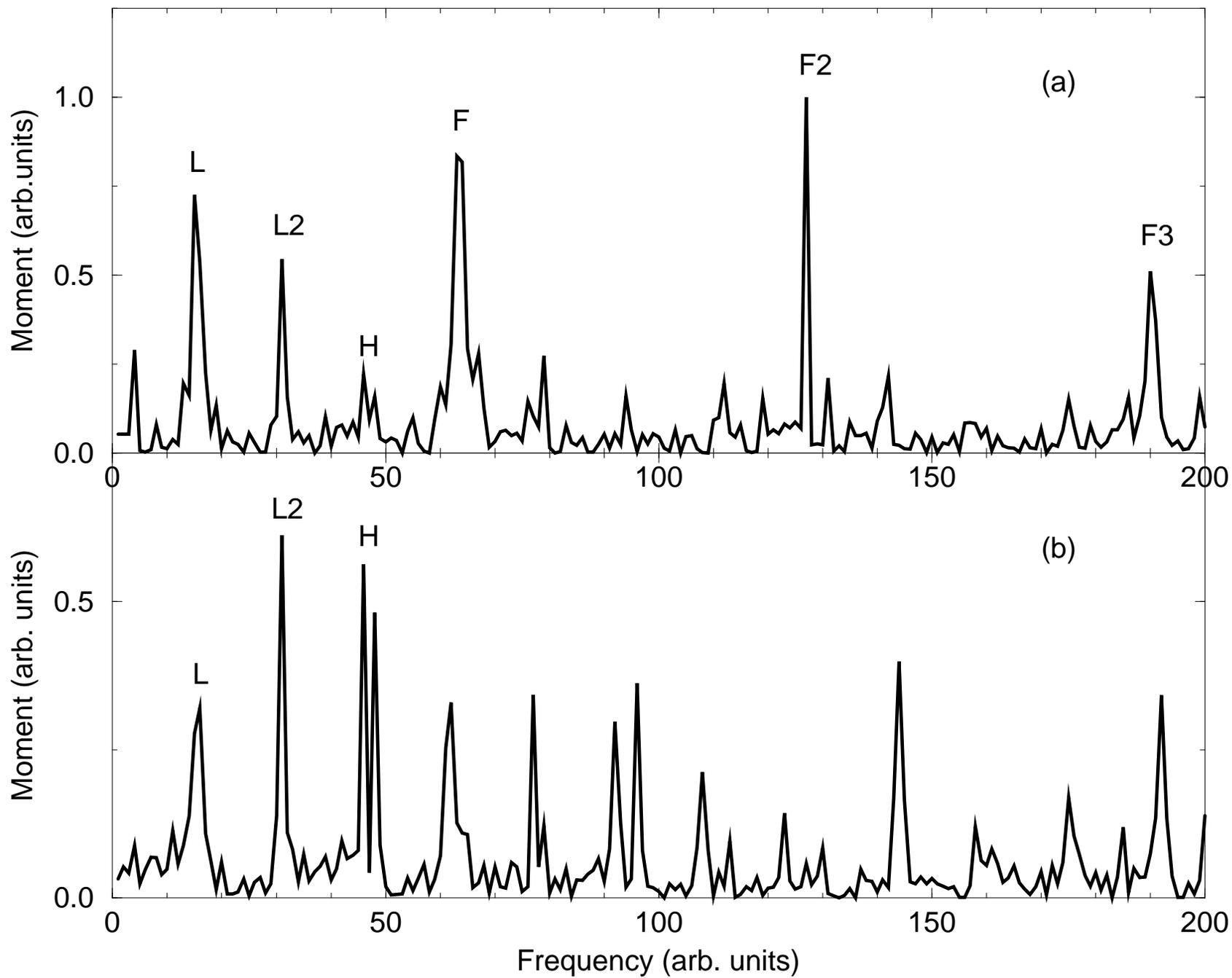
\end{document}